\documentclass[electronic]{vgtc}             

\ifpdf                                
  \pdfoutput=1\relax                   
  \usepackage{graphicx}                
  \DeclareGraphicsExtensions{.pdf,.png,.jpg,.jpeg} 
\else
  \usepackage{graphicx}    
  \DeclareGraphicsExtensions{.eps}  
\fi

\graphicspath{{figures/}{pictures/}{images/}{./}}

\usepackage{microtype}                 
\PassOptionsToPackage{warn}{textcomp}
\usepackage{textcomp} 
\usepackage{mathptmx}   
\usepackage{times}          
        
\usepackage{cite}                      
\usepackage{tabu}                      
\usepackage{booktabs} 
\usepackage{url}
\usepackage[hidelinks]{hyperref}
\usepackage[utf8]{inputenc}
\usepackage[small]{caption}
\usepackage{acronym}

\newcommand{\ie}            {\textit{i.e.}} 
\newcommand{\eg}            {\textit{e.g.}} 
\newcommand{\etc}           {\textit{etc.}} 
\newcommand{\rom}[1]        {\textsc{({\romannumeral #1})}} 
\newcommand{\folder}        {\url{https://cdv.dei.uc.pt/essys_sharing_uc}}
\newcommand{\site}          {\url{http://sharing-uc.dei.uc.pt}}
\newcommand{\esuc}          {\acs{essys}* Sharing \#UC}

\onlineid{0}

\vgtccategory{Research}

\vgtcinsertpkg
\newacro{gd}[GD]            {Graphic Design}
\newacro{nlp}[NLP]          {Natural Language Processing}
\newacro{ml}[ML]            {Machine Learning}
\newacro{essys}[ESSYS]      {Emotion Sonification System}
\newacro{midi}[MIDI]        {Musical Instrument Digital Interface}
\newacro{nlu}[NLU]          {Natural Language Understanding}
\newacro{api}[API]          {Application Programming Interface}
\newacro{url}[URL]          {Uniform Resource Locator}
\newacro{json}[JSON]        {JavaScript Object Notation}
\newacro{otvf}[OTVF]        {OpenType Variable Font}
\newacro{cc}[CC]            {Computational Creativity}
\newacro{ai}[AI]            {Artificial Intelligence}
\newacro{ac}[AC]            {Adaptive Computation}
\newacro{ec}[EC]            {Evolutionary Computation}
\newacro{iec}[IEC]          {Interactive Evolutionary Computation}
\newacro{json}[JSON]        {JavaScript Object Notation}
\newacro{rws}[RWS]          {Roulette Wheel Selection}
\newacro{iso}[ISO]          {International Organisation for Standardisation}
\newacro{rest}[REST]        {Representational State Transfer}
\newacro{css}[CSS]          {Cascading Style Sheets}
\newacro{uc}[UC]            {University of Coimbra}
\newacro{capc}[CAPC]        {Círculo de Artes Plásticas de Coimbra}

\title{\esuc: An Emotion-driven Audiovisual Installation}

\author{
S\'ergio M. Rebelo\thanks{e-mail: srebelo@dei.uc.pt} 
\and Mariana Sei\c{c}a\thanks{e-mail: marianac@dei.uc.pt} 
\and Pedro Martins 
\and Jo\~{a}o Bicker
\and Penousal Machado
}
\affiliation{\scriptsize University of Coimbra\\Centre for Informatics and Systems of the University of Coimbra\\Department of Informatics Engineering}

\teaser{
  \centering
  \includegraphics[width=\linewidth]{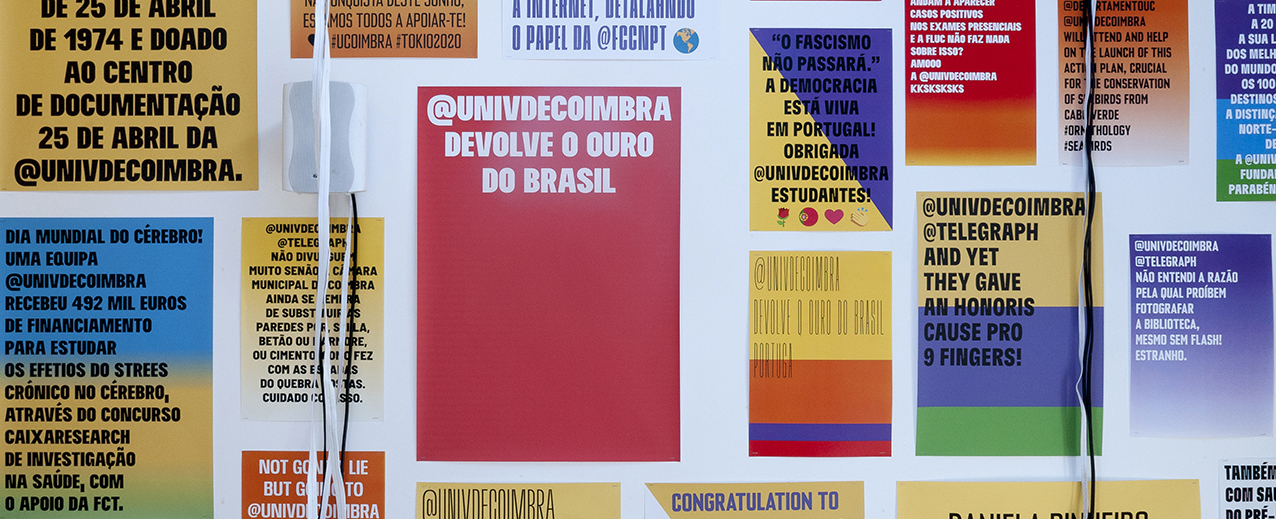}
  \caption{Detail photography of the \esuc~installation, displaying one of wall of installation crowed of posters generated based the emotion charge of its contents. Photography by Jorge das Neves.~\textcopyright~\acs{capc}, 2021.}
  \label{fig:teaser}
}

\abstract{
We present ESSYS* Sharing \#UC, an audiovisual installation artwork that reflects upon the emotional context related to the university and the city of Coimbra, based on the data shared about them on Twitter. The installation was presented in an urban art gallery of Círculo de Artes Plásticas de Coimbra during the summer and autumn of 2021. In the installation space, one may see a collection of typographic posters displaying the tweets and listening to an ever-changing ambient sound. The present audiovisuals are created by an autonomous computational creative approach, which employs a neural classifier to recognise the emotional context of a tweet and uses this resulting data as feedstock for the audiovisual generation. The installation’s space is designed to promote an approach and blend between the online and physical perceptions of the same location. We applied multiple experiments with the proposed approach to evaluate the capability and performance. Also, we conduct interview-based evaluation sessions to understand how the installation elements, especially poster designs, are experienced by people regarding diversity, expressiveness and possible employment in other commercial and social~scenarios.
}

\CCScatlist{
    \CCScatTwelve{Applied computing}{Arts and humanities}{Media arts}{};
    \CCScatTwelve{Information systems}{Information systems applications}{Multimedia information systems}{Multimedia content creation};
    \CCScatTwelve{Information systems}{Information retrieval}{Specialized information retrieval}{};
    \CCScatTwelve{Human-centered computing}{Visualization}{}{};
    \CCScatTwelve{Applied computing}{Arts and humanities}{Sound and music computing}{};
}

\begin{document}

\firstsection{Introduction}

\maketitle

The recent digitalisation of media and the consequent democratisation of access to digital technologies and the Internet have fostered the emergence of new Art and Design practices~\cite{boden2009a, paul2015a, richardson2016a}. These new practices are promoting a shift in the role of artists/designers, encouraging a creative practice where authors' roles are closer to curators than creators, \ie~authors build systems and tools that generate, with different levels of autonomy, artefacts instead of crafting themselves the artefacts~\cite{galanter2016a}. In this sense, the artistic use of procedural, generative and data-driven approaches is increasingly explored in the context of creative practices, especially by a new generation of people that is as much or more comfortable with digital media and tools than traditional media~\cite{tribe2007a}. 

In this paper, we present \esuc,~an audiovisual artistic installation that reflects upon contents shared online, on Twitter, about the university and the city of Coimbra. This artwork was commissioned by the Institute of Interdisciplinary Research of the \acf{uc}, in the context of the celebration of the 730 anniversary of \acs{uc}. The installation presented a hybrid nature, being accessible both in a physical art gallery and online. In both installation spaces, one may see a set of generated typographic poster designs, displaying tweets related to Coimbra, accompanied by an ever-changing ambient sound. The audience feedstock the audiovisual generation of the installation, influencing indirectly (through their tweets) the creation and the characteristics of the exhibited multimedia artefacts.

This installation was designed for the particular \emph{Museu} art gallery of \acf{capc}. This exhibition space was designed by the Portuguese artist Francisco Tropa and built in the context of \textit{anozero} 2017, the first edition of the Coimbra Biennial of Contemporary Art \cite{anozero2015a}. The space presents a participatory and easily accessible nature, being developed simultaneously as an urban sculpture and an always-open exhibition space.\footnote{One may read more about \emph{Museu} space at \url{https://2015.anozero-bienaldecoimbra.pt/19-museu/} (website visited on 2 June 2022).}\esuc~was exhibited in this space between July and December 2021. 
Figures~\ref{fig:teaser}~to~\ref{fig:physical-artwork-2} display the setup of the physical installation.

On the other hand, we also develop an online version of the installation, which is accessible at~\site. This version of the installation comprises an ever-changing audiovisual environment defined by the continuous happening of an autonomous \acf{cc} system. In this installation, this system is constantly generating audiovisual artefacts based on its perception of the emotional charge of the gathered tweets. 
Figure \ref{fig:online-artwork} presents a screenshot of the online installation. The online version of the installation was first developed to be part of the physical installation; however, we decided to make it publicly available to enable a safe visualisation due to the COVID-19 pandemic social distance restrictions at that time and, simultaneously, to allow the perpetual visit to this installation when the physical installation closes.

The main motivation behind this work was to produce an artwork that looks into how virtual data about a physical location could be apprehended and visualised by a virtual system in order to characterise the same physical location, through media that remits directly to this location (in this case, the typical posters' wall often observed in our everyday urban space). Furthermore, we had the intention of creating a space that enabled these online data to leave the borders of the digital space and blend itself with the corresponding physical space. In this sense, the proposed installation is designed to open a window of reflection on the meaning of what is a location nowadays, what are the variables and dimensions in this definition, and how the online data about a location could modify the perception and behaviours of its inhabitants and passers-by. Also, in the development of this artwork, we were interested in studying the potential advantages, disadvantages, and biases of the employment of autonomous \acs{cc} strategies to emotionally translate the data and, subsequently, generate creative artefacts based on it. 

The present audiovisual artefacts are designed by an autonomous \acs{cc} approach that performs emotion recognition in tweets and subsequently uses the results to inform the generation of audiovisuals exhibited on the installation. The proposed approach comprises three modules:
\rom{1} the \emph{Input Preprocessing} module, which employs a \acf{ml} approach to emotionally classify the tweets;
\rom{2} the \emph{Typesetter} module, which generates the poster designs; 
and \rom{3} the \emph{\acf{essys}} module, which generates the ambient audio developed based on the previous work of Sei\c{c}a \emph{et al.}~\cite{seica2017a}. 

We evaluate the outputs created by the proposed approach for understanding its communication and visualisation value from the audience's point of view, studying how the generated outputs are perceived by the audience. Moreover, we study the possible applicability of the present approach in other creative commercial scenarios. This way, we experimented with the proposed approach to study its performance and capability in generating poster designs, and to understand how people perceive these designs, both in expressive and creative terms. Thus, we generated outputs for several kinds of texts (with different lengths and textual and emotional purposes) to evaluate the module performance and capability for generating outputs. Furthermore, we conducted interview-based evaluation sessions with twelve participants to perceive the visual coherence, diversity, and emotional expressiveness of the outputs, the effect of audio on posters’ visualisation, and future applications of similar visual artefacts in other media.

The remainder of this paper is structured as follows. The next section comprises an overview of related work on the procedural generation of typographic layouts and emotional-driven musical pieces. Section 3 presents and describes our approach to designing the installation and the creative system behind that. Section 4 presents the evaluation experiments and the analysis of the results. Finally, Section 5 discusses our conclusions and points to future work.

\section{Related Work}
Since the advent of computers, in the 1950s, several artists, computer scientists, and creative practitioners had sporadically explored computational approaches and \acl{ai} techniques as the foundation for their artworks~\cite{boden2009a}. Nevertheless, in the last two decades, we observed the increasing employment of these approaches, especially because of the development of easier-to-use creative code environments~\cite{galanter2016a}. For instance, Paul~\cite{paul2015a} and Audry~\cite{audry2021a} present a solid overview of the field.

Recently, designers and artists also began to collect sentiment and emotional data (mostly online data from social media platforms) with the aim of developing audiovisual artefacts that reflect, in some way, the thoughts, emotions, and feelings present in these data. Harris and Kamvar \cite{kamvar2009a} developed \emph{We Feel Fine} by extracting emotional information from several blog entries around the Internet. The results were presented as one particle cloud of dots that could be filtered by varied criteria. Taraborelli \emph{et al.} \cite{taraborelli2010a} created organic tree-like structures that picture the online discussions about whether a topic should remain or be deleted from Wikipedia. Maçãs \emph{et al.} developed \emph{typEm} \cite{macas2019b}, a system to generate texts that compose the text, deforming the shape of glyphs based on the emotional values present in each sentence. Ong developed the installation \emph{\#home} \cite{ong2019a} where the tweet feed is translated into 3D forms that subsequently were 3D-printed on small scale and housed in a jar. Quintas and Sandoval developed the installation \emph{News Feed} \cite{quintas2019a} that generates audiovisuals based on the sentiment of news stories posted on the major worldwide online newspapers. Qin \emph{et al.} developed the bio-feedback system \emph{HeartBees} \cite{qui2020a} that translate a multi-dimensional emotion model into motion configurations that control the movements of a swarm. Also, they evaluated the perception of people of the generated swarm. The results showed that the people subjectively perceived the visualisations. Sonification has also been explored as an effective dimension for humanising approaches to data design. Examples range from physical manipulations that explore the relationship between sound and matter \cite{lenzi2018}, to communication of social issues with wider audiences with which the public can relate, as a multimodal experience in a website of Egyptian building collapses \cite{tactical2014, lenzi2020}, or the works of Brian Foo using generative art \cite{foo2012, lenzi2020}.

These artefacts explored the mapping of emotional data into different media and configurations. In this work, we explored the translation of these data into typographic composition and emotional-driven musical pieces. The following subsections comprehensively examine related works that directly address the technical scope of this work.

\subsection{Generating Typographic Compositions} 
Advances in digital technologies permitted new aesthetic possibilities, which conduct an innovative and flexible \acf{gd} practice~\cite{triggs2003a, paul2015a}. First explorations related to typographic layout generation were noticed already in the last half of the 20th century. We highlight the pioneering \emph{Plan for a Conceptual Art Book} (1971) of LeWitt~\cite{cramer2022a} as well as the work of Cooper and her students at the \emph{Visible Language Workshop} (\eg~\cite{cooper1989a, musgrave1996a}) and Maeda (\eg~\cite{maeda2000a}).

\acl{ec} approaches have been popular in creating layouts 
through both \acf{iec} (\eg~\cite{gatarski2002a}, \cite{onduygu2010a}, \cite{morcilllo2010a} or \cite{kitamura2011a}) and automatic fitness assignment strategies (\eg~\cite{goldenberg2002a}, \cite{purvis2003a}, \cite{lopes2022a}). Nevertheless, \acs{iec} strategies can provoke users' fatigue and inconsistent evaluations, and in automatic strategies are difficult for users to express their preferences to the systems~\cite{machado2008a}. Thus, there have been developed works blending automatic and interactive evolution (\eg~\cite{quiroz2009a} or \cite{rebelo2021a}); however, they are still time-consuming approaches and are not appropriate for a faster layout generation.

More recently, \acs{ml} methods have been also explored to support the layout generation taking into consideration the relation between elements and the learning of specific typesetting and design styles (\eg~\cite{zheng2019a}, \cite{li2021b}, \cite{guo2021a} or \cite{kikuchi2021a}). Nevertheless, these approaches require the use of annotated training datasets on the layout design. Currently, these datasets still include few data both in terms of quantity and diversity.

Template-based and suggesting frameworks are also a widespread approach to the generation of typographic layout designs. These approaches generate, suggest, and/or perform modifications in layouts based on the content (\eg~\cite{jacobs2004a} or \cite{venkata2011a}), the inter-relationships between elements on layouts (\eg~\cite{cleveland2010a} or \cite{donovan2014a}), or context-aware and/or external data (\eg~\cite{LUST2014a} or \cite{rebelo2019a}). Similarly, some procedural approaches allow the generation of layouts throughout a set of predefined bespoke data-driven or/and random procedures to define the features of a layout (\eg~\cite{gross2007a} or \cite{lafkon2011a}). Although these methods may present some limitations in terms of visual exploration, they generate consistent layouts in a time-efficient and effortless manner. 

There is also possible to identify some works that use sound input to influence the generation of typographic compositions. As far as we know, one of the first experiments in the \acs{gd} context was Maeda's \textit{Reactive Graphics}, where he considered the microphone as a type of input source to modify visuals (\eg~\textit{Reactive Square}~\cite{maeda1994a}). Subsequently, some authors used sound inputs to influence the visual features of their designs by varying the visual features of all elements on designs (\eg~\cite{givens2017a} or \cite{lopes2018a}) and/or modifying the letter shapes on their design (\eg~\cite{ali2010a} or \cite{parente2022a}).

\subsection{Generating Emotional-driven Music}
The use of computational approaches to generate musical compositions has been explored as a way to expand the boundaries of musical creations. In this line, algorithmic composition allows formal processes with lower or almost non-existent levels of human involvement to create music \cite{edwards}, using sets of predefined grammars or ones which evolve in time. The system \textit{MetaCompose} \cite{scirea} is an exploration of evolutionary music that may express distinct moods in a dynamic scenario, mixing a multitude of musical features, from harmony, melody, pitch, scale, and timbre to rhythms. Another example on the spectrum of rule-based systems is the work of Livingstone \textit{et al.}'s \cite{livingstone}, where they present a set of rules to build a musical score and its expressive abilities in a performance set. Using Russel's arousal-valence model \cite{russell}, it also varies a set of musical features, such as the tempo, loudness, or mode, in an attempt to induce certain emotional reactions in the public.
Another example focused on computer-aided composition is Cope's system that, based on recombination and mixing methods of known classical music works (of Bach and Beethoven, for example), allows the users to simulate new compositions by combining common patterns \cite{cope}. Eno, whose musical work focuses on exploring ambient music of environments, tackled the concept of \textit{process music} \cite{edwards} for composing his album \textit{Music for Airports}. The score is more of a set of graphical patterns which represent the system he created to make the music, exploring particular patterns of the instruments at half the intended speed and their timely repetitions \cite{eno}. Eno later explored self-generating musical systems, producing ever-changing ambient music compositions using sets of probabilistic rules, such as the \emph{Generative Music 1} album using \emph{SSEYO Koan Software} \cite{sseyo}, and his recent album \emph{Reflection}, available as an infinite piece through an iOS app \cite{eno}. 

\section{Approach}
\begin{figure}[hbt]
    \centering
    \includegraphics[width=\linewidth]{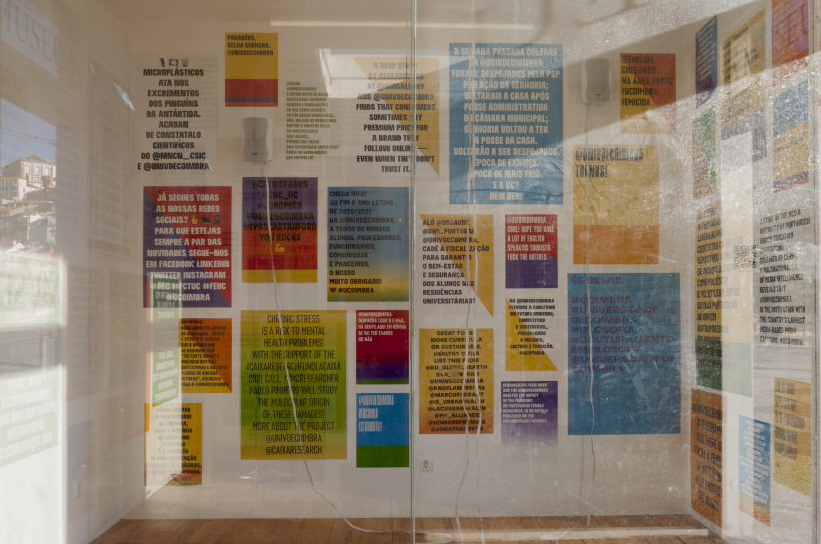}
    \caption{Photography of the physical installation displaying a wall crowded of posters. Photography by Jorge das Neves. \textcopyright~\acs{capc}, 2021.}
    \label{fig:physical-artwork}
\end{figure}

\begin{figure}[hbt]
    \centering
    \includegraphics[width=\linewidth]{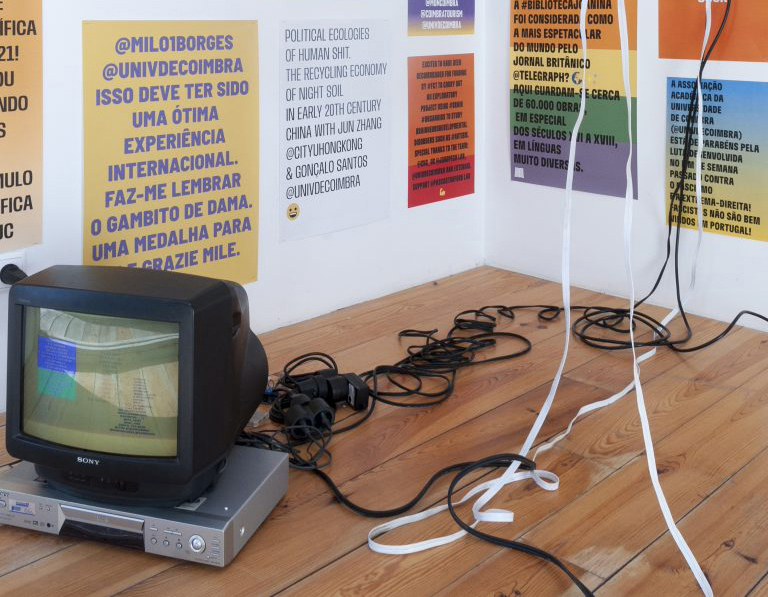}
    \caption{Photography of the physical installation displaying a wall crowded of posters as well as the television displaying the online installation running. Photography by Jorge das Neves. \textcopyright~\acs{capc}, 2021.}
    \label{fig:physical-artwork-2}
\end{figure}

\begin{figure}[hbt]
    \centering
    \includegraphics[width=1\linewidth]{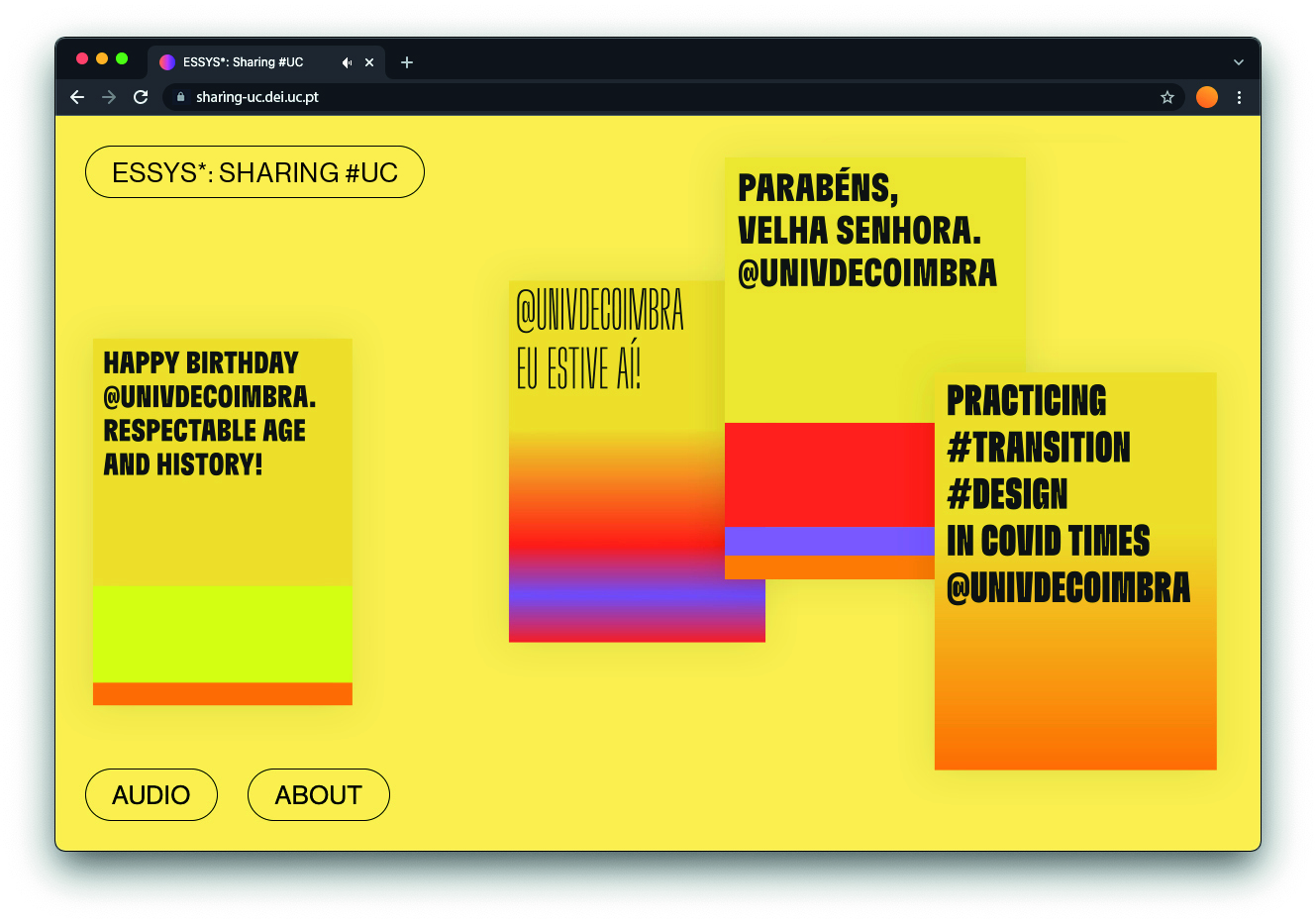}
    \caption{Screenshot of the online installation displaying posters generated by the present approach. The predominant emotion recognised in this environment was joy.}
    \label{fig:online-artwork}
\end{figure}

We present an installation composed of audiovisual elements that represent the content shared online on Twitter, regarding the city and the University of Coimbra. This installation is accessible both online and in a continuous open art gallery, in Coimbra's historic centre. This installation is aligned with our previous work on emotion-driven audiovisual generation~(see~\cite{seica2021a} and \cite{seica2017a}). The present installation is designed to promote the reflection about what is the meaning of location nowadays, creating a space where the online data about a location is blended and presented in the corresponding physical location. In the two installation configurations, one may observe a set of generative poster designs, concerning the content of gathered tweets, while listening to an ambient sound that reflects the emotional charge of the same tweets. The online installation, which is available at \site, presents an ever-changing audiovisual environment when posters are floating on the screen coupled with ambient audio (see Figure~\ref{fig:online-artwork}). Otherwise, the physical installation presents a collection of posters of multiple sizes and formats generated previously, together with one television displaying the online system running and serving to speakers the ambient audio (see Figures~\ref{fig:physical-artwork} and \ref{fig:physical-artwork-2}). 

The outputs are generated through the implementation of three autonomous modules: \rom{1} the \emph{Input Preprocessing}; \rom{2} the \emph{Typesetter}; and \rom{3} the \emph{\acs{essys}}.
The \emph{Input Preprocessing} module employs a neural \acl{nlp} approach to predict the emotional context of a tweet and splits the text into lines using a sentence tokeniser method. These data inform the outputs generated by the other modules, both visuals and audio. 
The \emph{Typesetter} module generates the poster designs informed by the gathered emotional context, through an autonomous greedy approach. 
On the other hand, the \acs{essys} generates \acs{midi} sonic compositions through a probabilistic rule-based approach also based on the gathered emotional context. 

\begin{figure*}[tb]
    \centering
    \includegraphics[width=\textwidth]{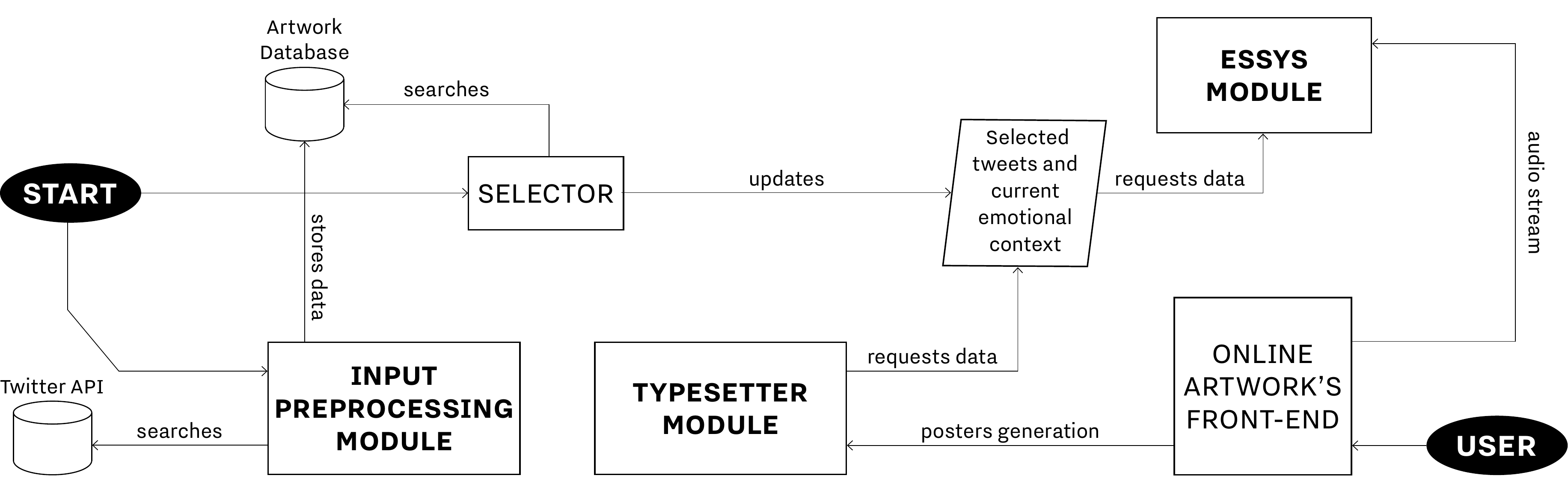}
    \caption{Installation support system overview.}
    \label{fig:scheme}
\end{figure*}

We developed a web-based system to support the installation. The system is continuously searching and collecting the related tweets, via Twitter \acs{api}, preprocessing and analysing them, and, finally, storing the tweets and analysis data in a database. In parallel, it selects the tweets to feed the audiovisual generation. This way, 10 times per second, the system queries the database and selects 20 tweets based on their newness. Next, a \acf{rws} method selects one tweet based on the date and time that they are created. More recent tweets have more probability of being selected. Subsequently, it assigns to a tweet one random lifespan between 1 and 3 minutes. Due to the space available on the regular computer screens, we defined empirically that the installation could only exhibit five posters simultaneously. Thus, when the maximum number of posters is in the exhibition, the selection is paused until a poster is removed. When a user accesses the website, the front-end side of the system requests information about the tweets selected, generates and presents the posters, and triggers the emotional-related audio stream produced by~\acs{essys}. Also, one by second, it consults what are the currently selected tweets. Whenever a tweet is added and/or removed from this list, it handles updates on the poster visualisation and audio. To simplify the connection between the system parts, we developed a \acs{rest} \acs{api}. Figure \ref{fig:scheme} presents a blueprint of the presented web-based system.

\begin{figure}[t]
    \centering
    \includegraphics[width=1\linewidth]{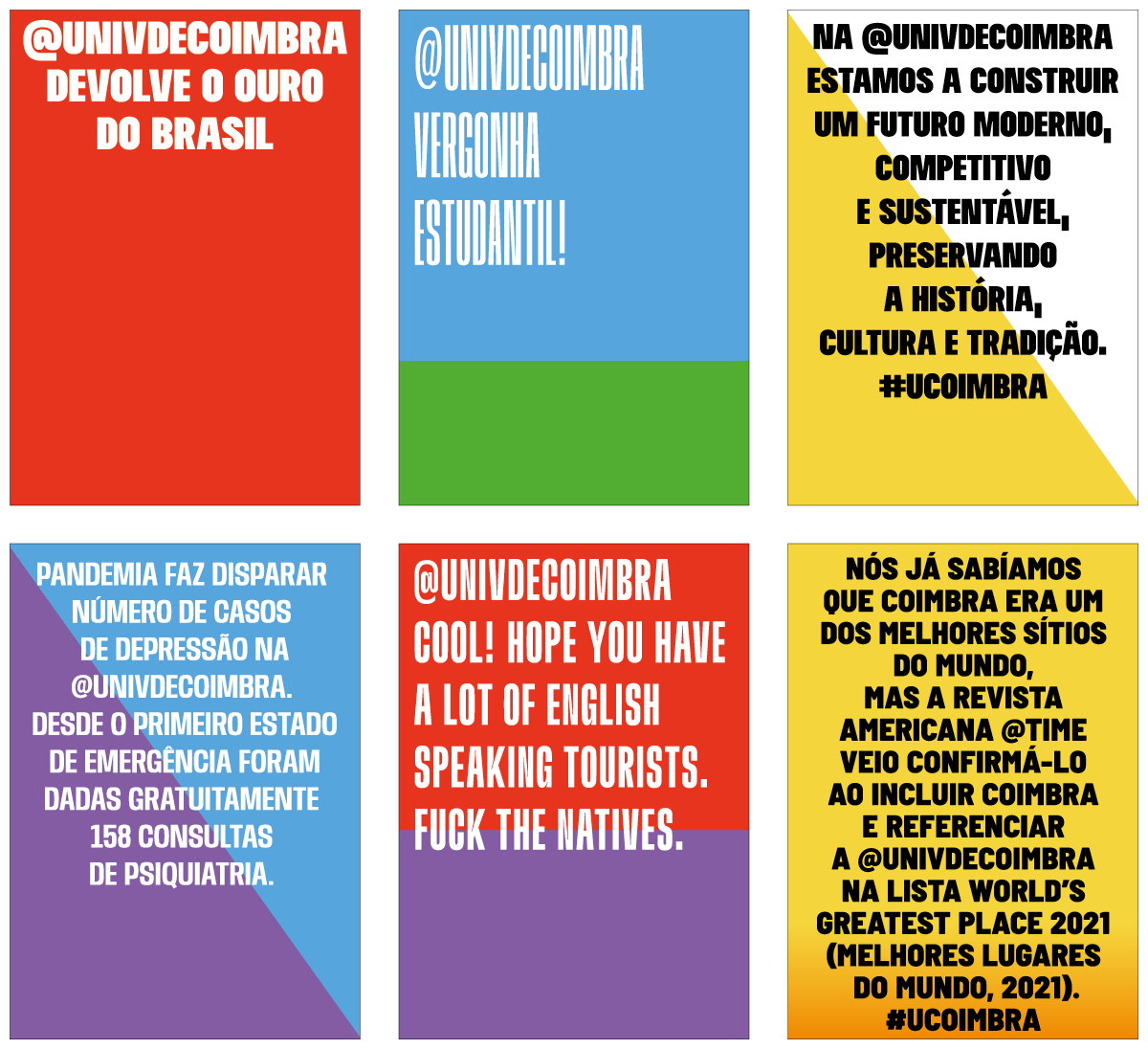}
    \caption{Some typical generated poster designs presented on the installation, using contents about the \acs{uc}. Supplementary documentation and demonstration videos are available at~\folder.}
    \label{fig:exp-installation}
\end{figure}

The presented framework uses style configuration files to inform their generation process. In these files (\acs{json} files), the possible attributes and features of the outputs are defined, such as the relationship between visual features, \ie~colours (colour configuration file) and typefaces (typeface configuration file), and classification results. This way, one may effortlessly reconfigure the system behaviour, adjusting the outputs to other contexts, \eg~fine-tuning the visual features or relating them with other classification results. Figure \ref{fig:exp-installation} portrays some posters displayed in the installation. Supplementary documentation, demonstration videos, and the utilised configuration files are made available at~\folder.

\subsection{Input Preprocessing module}
\label{sec:inputModule}
The \emph{Input Preprocessing} module processes the tweets to recognise the emotions embedded and divide them into lines. In the end, the resulting analysis data is stored to be accessible by other modules. 
The tweets are gathered, querying the Twitter \acs{api}, by tweets posted with popular hashtags related to the city of Coimbra, as well as tweets posted by \acs{uc} and Coimbra local and regional administration~accounts.

We trained a \acl{nlp} neural network to understand the emotional intensity of a tweet, using a corpus built based on the emotion-intensity annotated multi-label tweets dataset of~Mohammad \textit{et al.}~\cite{mohammad2018a}. This dataset comprises labelled tweets, in the English language, classified as neutral or as transmitting one, or more, of eleven emotions that best represent the mental status of the tweeter. We focused the system on recognising the eight emotions of Plutchik’s theoretical model~\cite{plutchik1980a}: anger; anticipation; disgust; fear; joy; sadness; surprise; and trust. 

The analysis begins with preprocessing of the tweet's text for the classification. First, the text is translated to English, if it is not already in that language, using the web cloud \acs{api} IBM Watson Language Translator V3\footnote{One may read more about IBM Watson Language Translator \acs{api} at \url{https://www.ibm.com/cloud/watson-language-translator} (website visited on 25 May 2022).}. 
After, the system collects some features related to text, \ie~user's information, retweet data, mentions to users, \acs{url}s, hashtags, and the used emojis (and their significance).
Next, the tweet is classified by the neural network, recognising the related emotions and their intensity scores. Finally, the predominant emotions are determined based on a threshold score.

This module is also responsible for subdividing the text into lines. We employ a \emph{Sentence Boundary Detection}~\cite{reynar1997a} algorithm to divide the text into sentences. Then, the resulting long sentences are stochastically subdivided according to a predefined optimal range of words and the pagination tradition~\cite{bringhurst2004a, lupton2010a}, \ie~minimising the production of typographic widows, orphans as well as the place of small words at the end of lines. 

\subsection{Typesetter Module}
The \emph{Typesetter} module is responsible for the generation and rendering of poster designs. To do that, this module implements \rom{1} the \emph{styling} method (which defines the visual features of posters) and \rom{2} the \emph{typesetting} method (which composes the text).

\subsubsection{Styling method} 
\begin{figure*}[tb]
    \centering
    \includegraphics[width=\textwidth]{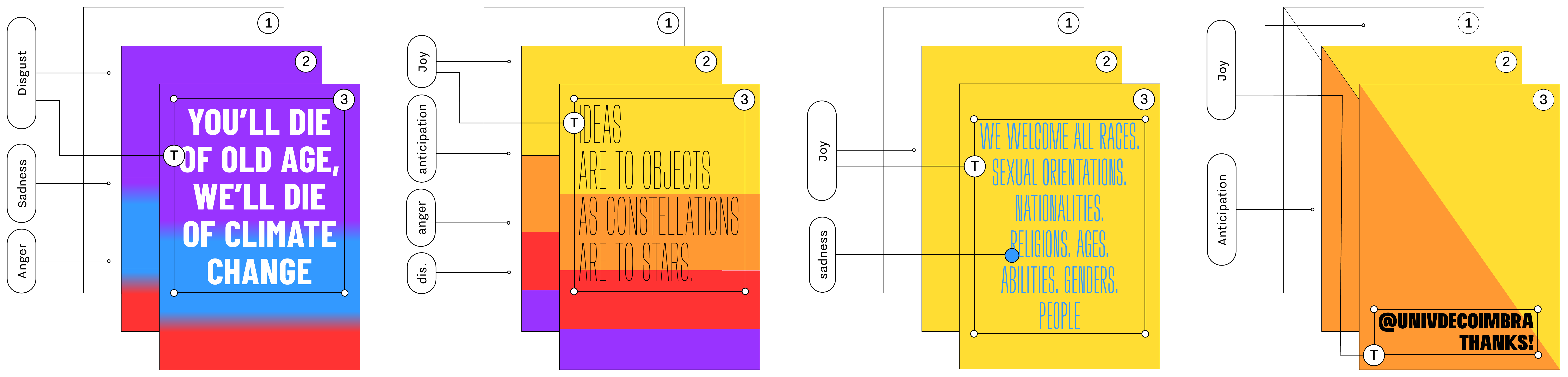}
    \caption{Schematic illustrates the architecture of some typesetting module example outputs, unveiling the impact of emotions recognised in the variation of background, chromatic and typographic styles (T) of generated designs. In this picture, emotion names are abbreviated, such as ANG (anger), ANT (anticipation), D (disgust), J (joy), and S (sadness). The recognised emotions are organised by classification's confidence from top to bottom.}
    \label{fig:legend-outputs}
\end{figure*}

The visual features of posters are defined by taking into account the emotional data previously recognised and the predefined data in configuration files. Figure~\ref{fig:legend-outputs} illustrates the correlation between the recognised emotional data and the background, chromatic and typographic styles of generated poster designs.

The poster format is randomly selected based on the regular \acs{iso} standards formats (A, B and C). We defined the colour-emotions relation based on the emotion wheel of Plutchik \cite{plutchik2001a}. We loaded eight \acf{otvf} into the system. We empirical defined their emotional relationship based on the works of Koch~\cite{koch2012a} and Hyndman~\cite{hyndman2016a}. The selected typefaces were chosen based on the possibility of variation of both font-weight and character width (or font-stretch) axes. We preferred typefaces that styles are traditionally associated with poster aesthetics and those we observed that are visually related to the present background styles. This way, we favoured sans serif letterforms over serif ones since they are aligned with our aesthetic preferences for the outputs of this installation.

The posters' background style is selected using an \acs{rws} method informed by the classification results and the selection probability of each background style to the recognised emotions. There are available four background styles: \rom{1} \emph{solid;} \rom{2} \emph{diagonally halved;} \rom{3} \emph{solid divided;} and \rom{4} \emph{gradient}. 
\emph{Solid divided} style can only be selected when two or more emotions are recognised. The other styles can be selected for any classification results. 
When no emotion is recognised in the tweet (\ie~neutral), the text is composed in black over a white background. These background styles are specified in a specific namespace (or package), enabling further development of new ones. 

In these background styles, the employed colours are chosen based on the emotions recognised in the text. Thus, the colours are chosen by an \acs{rws} method considering the predefined set of colours and their selection probabilities by emotion. Neutral colours are restricted to black and white. We also developed a legibility checker that verifies whether the selected colours (both in typography and background) fulfil a set of requisites and do not compromise the posters' message legibility. 

The \emph{solid} style fills the background with one solid colour. First, the background colour is selected amongst the set of colours related to the predominant emotion, plus white. The selection probability of white colour was empirically defined at 10\%. Afterwards, it selects the typography colour. When the background colour is selected as white, it is no longer a colour option for the typography colour, which is then selected in a way similar to the background from a predefined set. Otherwise, it chooses between using a neutral or a colour related to the second most predominant emotion. The more similar the score between two predominant emotions, the higher the probability of selecting a non-neutral colour. When no secondary emotion is recognised, typography colour is neutral. 

The \emph{diagonally halved} style designs and fills a two triangle diagonal composition. This way, it randomly selects one triangle and fills it with a colour related to the predominant emotion. Next, it chooses between filling the other triangle using white, or a colour related to the second most predominant emotion. In this style, also the more similar the score of the two predominant emotions, the higher the probability of selecting an emotion-related colour. When only one emotion is recognised, the second triangle is coloured white.

The \emph{solid divided} style creates a composition of rectangles that conveys the predominant emotion in the tweet. This way, the height of rectangles represents the emotion classification scores, directly translated into the percentages of the poster's height. The rectangles' fill is defined based on the recognised emotions in the same way that is presented in other styles. Typography colour is randomly defined in a neutral colour that does not compromise the poster's legibility.

The \emph{gradient} style is comparable to the \emph{solid divided}; however, instead of filling rectangles with solid colour, it uses linear colour gradients. It also employs a full-height rectangle coloured with a gradient of the selected colour to white, whenever only one emotion is recognised. The position of the gradient end-point (the white one) is placed in the last vertical quarter of the poster, mapped according to the emotion classification score, \ie~the higher score, the near the bottom edge is the end of the gradient.

The poster's typeface is also chosen based on the emotional context of the tweet. It also employs an \acs{rws} method that selects one typeface based on the predominant emotion and the predefined selection probabilities. In the end, the text alignment (\ie~right, centre, or left) and text box alignment (\ie~top, middle, or button) are randomly determined.

\subsubsection{Typesetting method} 
The \emph{Typesetting} method composes the posters, using an iterative greedy approach to modify the visual features of an \acs{otvf} with multiple axes.

The typesetting procedure begins by creating a text box of the same size as the poster, divided into a grid with one column and the same number of rows as the computed line division. 
The text is then placed on the text box according to the grid. The text leading is defined as the height of the rows. The text size is computed based on the leading, looking at a predefined relationship between the text leading and size for the selected typeface.
Thus, we ensure that this method performs the traditional leading adjusts based on the size.
Subsequently, it verifies if all the text lines are rendered inside of the poster; if so, the generation is finished. Otherwise, it randomly selects an operator to change the visual features of the employed typeface. To do that, we developed two operators: \rom{1} \emph{size modifier;} and \rom{2} \emph{axis modifier.} Both operators change the visual features of all text lines. When the selected typeface has no variation axes, it only employs the \emph{size modifier}.
This last step is repeated until the whole text lines are rendered within the poster, or a too high number of attempts is achieved. Video demonstrations of this procedure are available at~\folder.

The \emph{size modifier} operator decreases the typeface size at a certain value. This value is computed based on a predefined range for the used typeface, which increases along with the number of attempts,~\ie~the more attempts to compose the poster, the higher the decrement. When this operator is employed, the height of rows in the grid is defined according to the new text size and the grid margins increase based on the present text box alignment. This operator could be employed until a minimum row height is reached. 

The \emph{axis modifier} operator changes the value on a typeface's axis. By default, the operator modifies two axes: font-stretch; and font-weight. However, it can also handle other axes when defined in the configuration file. The axis to be modified is randomly chosen among the available options. The font-stretch axis can range between a maximum (default as ultra-expanded, \ie~200\%) and minimum range (default as extra-condensed, \ie~50\%). When this axis is modified, its value decreases in a \acs{css} keyword value (\ie~around 16.6\%). The font-weight axis can range between 950 (Extra Black) and 100 (Thin). Whenever this axis is modified, its value decreases by 10. The axes' values are reset to their defaults once their values reach a minimum and the size is changed, at least four times, since the last modification of any axis.

\subsection{\acs{essys} Module}
The \emph{\acs{essys}} module generates the ambient sound that reflects the predominant emotion on the posters presented on installation. This sonic composition is generated by a rule-based system where the musical parameters are dynamically chosen according to predefined probabilities, which depend on the emotion being portrayed. 

The audio generation is guided by two major musical aspects: melody and harmony. The melody is understood as a set of notes whose type (scale note, chord note or chromatism) and duration (from whole to eighth notes) are parameterised, as well as the intervals between them that shape the leaps of the melodic line. Each note is sequentially chosen depending on the probabilities of these parameters, guided by a melodic scale that is, in turn, defined by the harmonic progression chosen for each emotional context. These progressions are predefined for each emotion. Finally, multiple timbres of mostly synthesised sounds were chosen for each emotional context. This module is developed based on the work of Sei\c{c}a \emph{et al}~\cite{seica2017a}, adapting the timbres of synthesised sounds to use free VST plugins, more specifically the Spitfire Audio LABS4 virtual instruments.

\section{Evaluation}
We conducted two experiments with the presented approach to study and analyse how its outputs are perceived from a communication and visualisation perspective, as well as the generative and exploratory capability of the \emph{typesetter} module as a tool for \acs{gd} exploration. 
In the first experiment, we assessed the generative capability and performance of the \emph{typesetter} module.
In another experiment, we conducted interview sessions to study how the installation outputs (both poster designs and ambient audio) are experienced by people, in terms of diversity, expressiveness and possible employment in other commercial and social scenarios.

\subsection{Performance Evaluation}
\begin{figure}[hbt]
    \centering
    \includegraphics[width=1\linewidth]{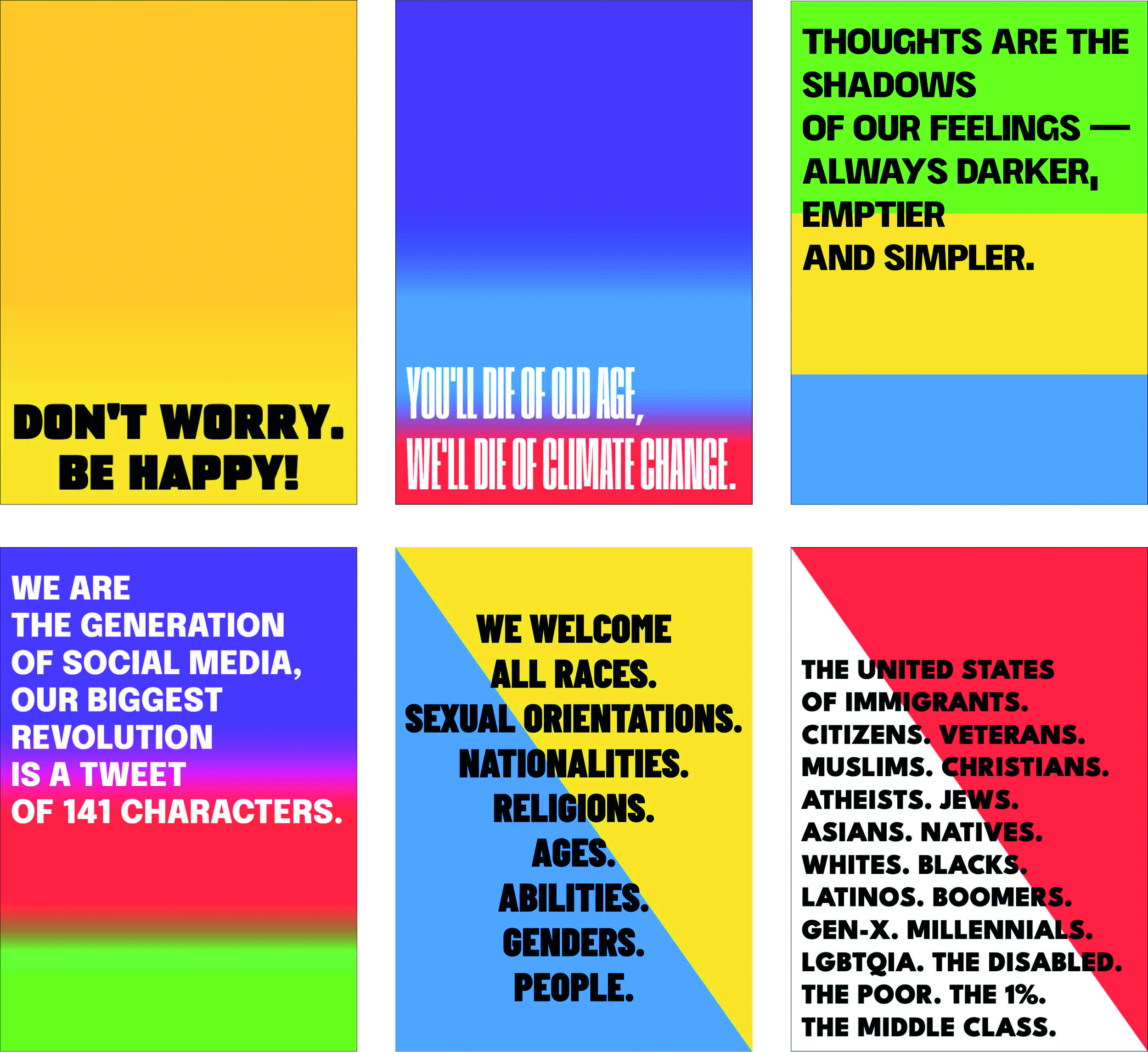}
    \caption{Some typical poster designs generated by the present approach during the evaluation stage. More results are available at \folder}
    \label{fig:results}
\end{figure}

We experimented with the \emph{typesetting} module to study its capability and performance for generating poster designs. We selected 16 texts with varied lengths and multiple textual and emotional purposes. The length of these texts ranges between 22 and 219 characters, regarding almost all the available sizes on Twitter. We used the \emph{Input Preprocessing} module to divide the text into lines as well as determine the emotional context of the tweets. We generated 300 posters for each text. Figure \ref{fig:results} displays some of the obtained outputs. 

This experiment unveiled that the module designs posters consistently presented a finished and legible poster design in all the evaluation runs. Also, we observed that it can generate poster designs in few operations (mean ($\bar{x}$): 61.25, median ($\widetilde{x}$): 31.50) and, consequently, time ($\bar{x}$: 0.88'', $\widetilde{x}$: 0.27''), in a standard web browser.

\begin{figure}[hbt]
    \centering
    \includegraphics[width=1\linewidth]{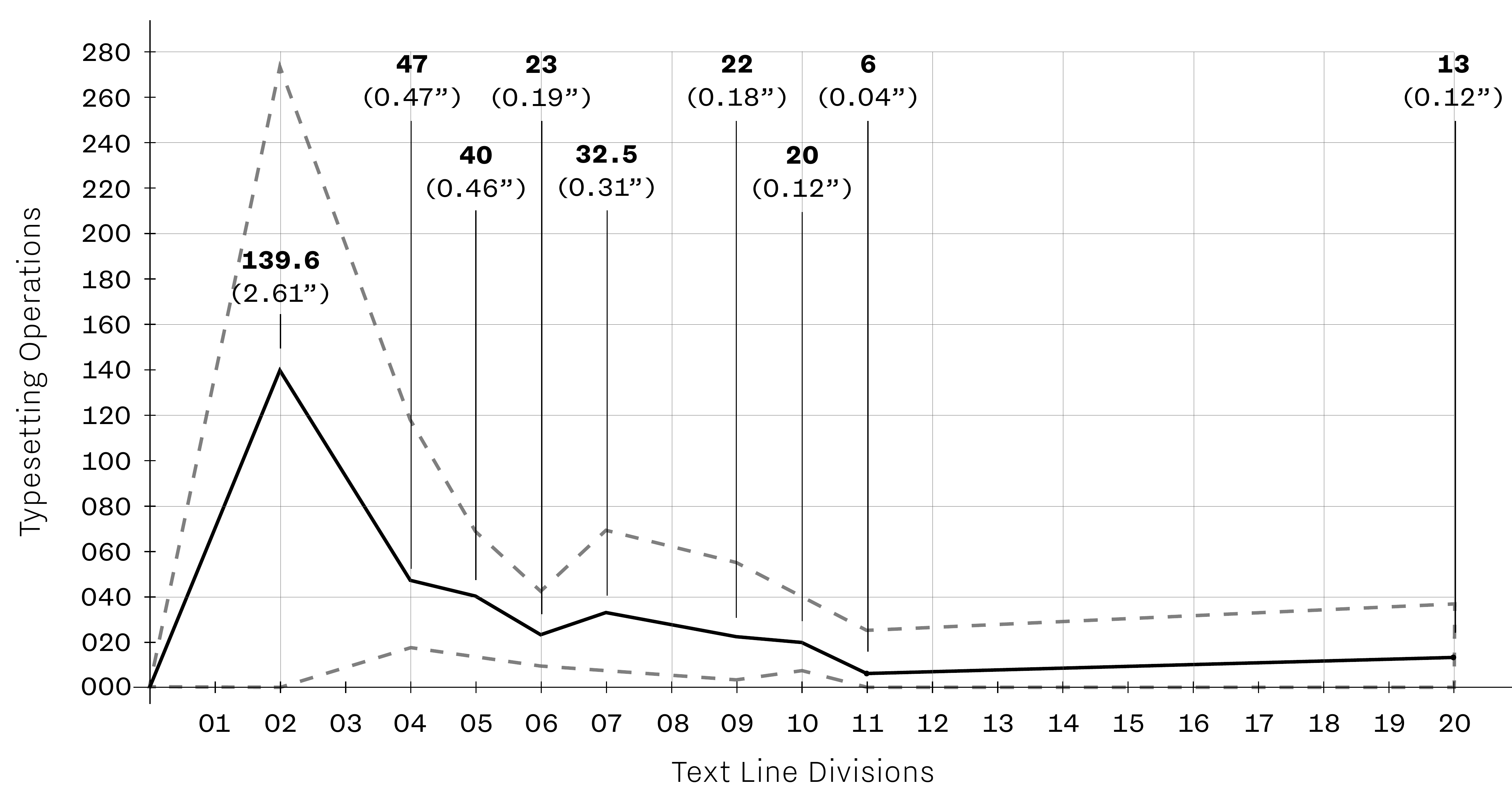}
    \caption{Performance evaluation results according to the number of line divisions (x-axis) and the number of typesetting operations required to generate a finished and legible poster design (y-axis). The solid line represents the average number of attempts by line divisions. The dashed lines delimit the range of needed attempts under the same condition.}
    \label{fig:exp1-plot}
\end{figure}

We observed that the quantity of operation is related to the number of line divisions, rather than the number of characters or words. In Figure \ref{fig:exp1-plot}, we can observe the relation between the number of typesetting attempts required to generate a poster and the number of line divisions. One can see that posters are generated faster when the number of line divisions is higher. When the text is divided into only two lines, on average, there are need 139.6 operations to generate a poster; nevertheless, the generation is more irregular, \ie~the possible range of needed operations is higher and varies between 0 and 272 operations. The generation becomes more regular, in terms of the number of operations when the number of lines increases. We also note that the typesetter greedy method favours the composition with narrower fonts because these fonts save space and allow composing the lengthier text lines on posters.

\subsection{Interview Evaluation Sessions}
We conducted interview evaluation sessions to understand how the installation’s outputs are perceived by viewers, concerning their visual diversity and coherence and the predefined emotional mapping. We were also interested in understanding the potential use in other commercial and social scenarios and their implications. In these sessions, we presented several posters (sometimes alongside an audio composition) and requested people to answer some quantitative questions and, following, document their perceptions and justify their answers. In total, we interviewed twelve participants. The age group of the participants ranged from 23 to 31 years old. Half of the participants (6) were professional graphic designers (G1), whose work experience ranged between two and eight years. This way, we ensured professional perspectives in the study of the potential of this approach, or other like, in commercial scenarios. The other half (G2) were from varied backgrounds. Table \ref{tab:participants-sumary}~further details the~participants.

\begin{table}[tb]
  \caption{Profiles of the participants for the interview evaluation sessions. In the graphic designer's participants, we present the number of years of professional experience of each participant along with the background (between parentheses).}
  \label{tab:participants-sumary}
  \scriptsize%
	\centering%
  \begin{tabu}{l c l c l}
  \toprule
   Participant & Age & Background & Gender \\
  \midrule
	P1 & 28 & Graphic Designer (06) & F \\
	P2 & 26 & Graphic Designer (03) & F \\
	P3 & 29 & Nurse & F \\
	P4 & 31 & Graphic Designer (08) & M \\
	P5 & 23 & Unemployed & M \\
	P6 & 28 & Graphic Designer (04) & M \\
	P7 & 26 & Graphic Designer (03) & M \\
	P8 & 30 & Computer Science Researcher & M \\
	P9 & 31 & Architect & M \\
	P10 & 26 & Graphic Designer (02) & M \\
	P11 & 23 & Unemployed & F \\
	P12 & 28 & Software Developer & M \\
  \bottomrule
  \end{tabu}%
\end{table}

The interviews were divided into four stages. The first (S1) focused on the visualisation of the posters' collections, followed by the classification of their visual coherence, on a scale between 1 (not cohesive) and 5 (very cohesive), and a justification. The first collection presented posters designed with the same typeface and distinct background styles. The next four collections were composed of posters designed with the same typeface and background styles. The last one presented poster designs with distinct typefaces and background styles. 

In the second stage (S2), we displayed seven posters, one at a time, and asked participants to define and justify the emotions that they perceived during the visualisation of each poster. In the following stage (S3), we showed five posters, one at a time, alongside an audio sample generated based on the same emotional context, requesting the participants to complete the same task as in S2. Two of the posters presented were also displayed in S2. In the end, we asked about the importance of audio in the perception of the posters' message. In the last stage (S4), we questioned participants about the potential and implications of the presented approach, or similar ones, in their daily basis consumption of information and social contexts. We extended the question to commercial and professional contexts if the participant was a graphic designer.

Concerning the visual diversity and coherence of posters (S1), the evaluation sessions proved the \emph{typesetter} module generates diversified poster designs; however, it was revealed to be more efficient in creating visually cohesive outputs. We observed that the perception of visual diversity is more related to background style than typeface style. All interviewed considered that the relation between the used colours, in the set of posters, is the decisive factor to create visual diversity or coherence. They mentioned that it was possible to identify a chromatic connection even when the visualised posters are designed with distinct backgrounds and typography styles. Nevertheless, 66.66\% (4) graphic designers (G1) and 50\% (3) of other participants (G2) referred to those posters with distinct backgrounds and typography are diverse enough, to the point that they could be designed for different contexts and/or by various designers. The background style that creates the more coherent output is \emph{halved diagonally}, mostly because of the easy-to-recognise triangle composition.  Every participant considered the collection of posters with this background style it was visual cohesive and most of both types of participants (G1:4, G2:4) even considered them extremely cohesive. Otherwise, most of the participants (G1:4, G2:3) considered that the posters with \emph{solid} background style allow the generation of more diverse outputs, mostly because this style does not present a distinguishing feature set. Finally, these sessions unveiled that text and text box alignment did not influence the perception of diversity.

The definition of the emotion charge is a subjective task influenced by the social and cultural context of the observer. The results of the S2 interview stage showed this, often unveiling contradictory answers for the same poster. Even when two people perceive the same emotion, they often justify it with varied reasons. Over the interviews, people supported their answers based on their examination of the posters' features, such as the visual style, the typeface, the colour pallet, the message,~\etc~Nevertheless, half or more of the interviewed people identified in most of the posters, at least, one emotion which was also identified by the neural classification of the \emph{Input Preprocessing} module. We observe that the recognition values were higher in the graphic designers' group (6 posters) than in the remaining participants (4 posters). Joy was the emotion more consensual recognised, being identified by almost all kinds of participants in two posters where it is one of the predominant emotions, 75\% (G1:4, G2:5) and 83.33\% (G1:5, G2:5), respectively. Anger and sadness were also recognised by, at least, 50\% of the participants, being more recognised by graphic designers. Disgust was the least recognised emotion, being only identified by one participant of each category (16.67\%). 

We obtained comparable results when posters were visualised along with an audio sample generated based on the same emotional context (S3). In the two posters already presented in S2, the recognition values stayed similar in the two categories of participants. All participants considered that the audio enhance the posters' visualisation, creating a context for their perception and, consequently, emphasising (or compressing) their message. However, some participants, especially non-designers participants (G1:2, G2:4), referred that it also can create confusion when the viewer does not understand the connection between the audio and the poster.

Concerning the current and future potential and implications of the presented approach, or similar ones, in (S4) most of the participants (G1:5, G2:3), particularly graphic designers, directly referred that the presented approach can be used to automatically highlight some contents in social networking and/or digital media, due to their eye-catching visual features. Some participants (G1:2, G2:3) also stated these kinds of audiovisuals, due to their automatic and emotion-driven nature, could aid in recreating the context of the author of the message, creating a context to better understand the reading message. However, in both scenarios, there is recognised that the massive employment of these approaches could have an opposite effect. All graphic designers identified that these kinds of approaches present a co-creative potential, allowing designers to create a framework (\eg~design system or visual identity) that everyone could easily use to create new designs. Most of the designers (4) also referred that this approach has exploratory potential, enabling the designer effortlessly to explore visually and conceptually new~ideas.

\section{Discussion and Conclusion}
We presented \esuc, an audiovisual installation artwork that combines the exhibition of generative posters and ambient audio. The audiovisuals presented in the installation are generated automatically by a \acs{cc} approach, considering the emotional context about the \acs{uc} and the city of Coimbra, based on the tweets about it. In this sense, the installation artefacts are defined indirectly by people, through their interaction with the related subject on cyberspace, \ie~on Twitter. This installation was commissioned by the Institute of Interdisciplinary Research of the \acs{uc}, in the context of the celebration of the 730 anniversary of \acs{uc}.

The present installation presents a hybrid nature being accessible both in a physical art gallery and online. The physical installation was presented, between July and December 2021, in a particular continuous open \textit{Museu} art gallery of \acs{capc}. On the other hand, the online installation is an ever-changing online environment where one may assist the installation based on the live generation of poster designs and ambient sound based on the perception of the present creative system.

In this artwork, we were motivated to create a space that allows the audience to visualise the data available online about a location, in the corresponding physical location. The proposed approach transforms these online data into audiovisual artefacts to promote the perception and visualisation of the same data, in the physical location, in media traditionally related to physical spaces. This way, this installation creates one space where the digital and the physical dimensions of a location blend between them. Moreover, the proposed installation created a window of reflection on what is the extension of location nowadays. Its exhibition space was designed to promote the reflection on how a physical space could be mirrored, visualised, perceived and defined online, promoting the audience's thoughts about the impact that online data currently have on the definition of the characteristics of a location and/or in the change of behaviour its inhabitants and passers-by. 

The generated audiovisual artefacts are designed by an autonomous \acs{cc} approach based on three modules: \rom{1} the \emph{Input Preprocessing} module, which performs the emotion recognition on the tweets; \rom{2} the \emph{Typesetter} module, which generates typographic posters using an iterative greedy approach; and \rom{3} the \emph{\acs{essys}} module, which generates the ambient audio. In the context of this work, we also analysed the use of autonomous \acs{cc} approaches to emotionally translate the data and, subsequently, generate creative typographic and auditory artefacts.

We performed evaluation experiments to study and analyse the perception of generated outputs and the audiovisual environment. Also, we were interested in the generative and exploratory capability of the present approach. In this sense, we conducted a performance evaluation to assess the generative capability and performance of the \emph{typesetter} module. Also, we conducted interview sessions with people, including \acs{gd} professionals, to understand how installation outputs are perceived by people, focusing on terms of diversity, expressiveness, and possible employment in other creative scenarios. These evaluation sessions unveiled that the proposed approach generates posters considered diverse and/or coherent by designers and non-designers in a time-efficiently manner. 

The emotional perception of one audiovisual is a subjective task influenced by the social and cultural background of the person who experiences it. However, during our interview sessions, we observed that some emotions recognised by the neural classifier (especially joy, anger, and sadness) are often perceived by participants in the posters, accompanied with or without the audio component. Interview participants consensual recognise that the audio dimension enhances the posters' visualisation, but some recognise that it can also lead to confusion.

Since the current installation was presented in the public space, it allowed to publicly present a set of subjects and matters regarding the city of Coimbra that, although often discussed online, rarely have been transposed to the city's urban space. This way, this installation was not only a showcase of the most common online Coimbra-related discussions, but it also provided one dedicated site to present them in the corresponding urban space. For instance, these topics include the Portuguese colonial inheritance and its legacy in Coimbra city and university, the current housing and gentrification issues in Coimbra's Historical Centre, the regional perception of the present national and worldwide political circumstances, the consequence of the COVID-19 pandemic, the national and worldwide influence of research institutes present on Coimbra,~\etc~The presented audiovisual outputs were created indirectly by the audience, based on the users' participation online. They expose multiple (often opposite) personal points of view about the same subject. In this sense, the installation's space was also a collective visualisation of the city, overviewing the predominant thoughts about it. Also, over time, the installation outputs, when observed time-related and collectively, are an archive of how people (both inhabitants and non-inhabitants) see and feel the city in its multitude of dimensions.

\begin{figure}[hbt]
    \centering
    \includegraphics[width=1\linewidth]{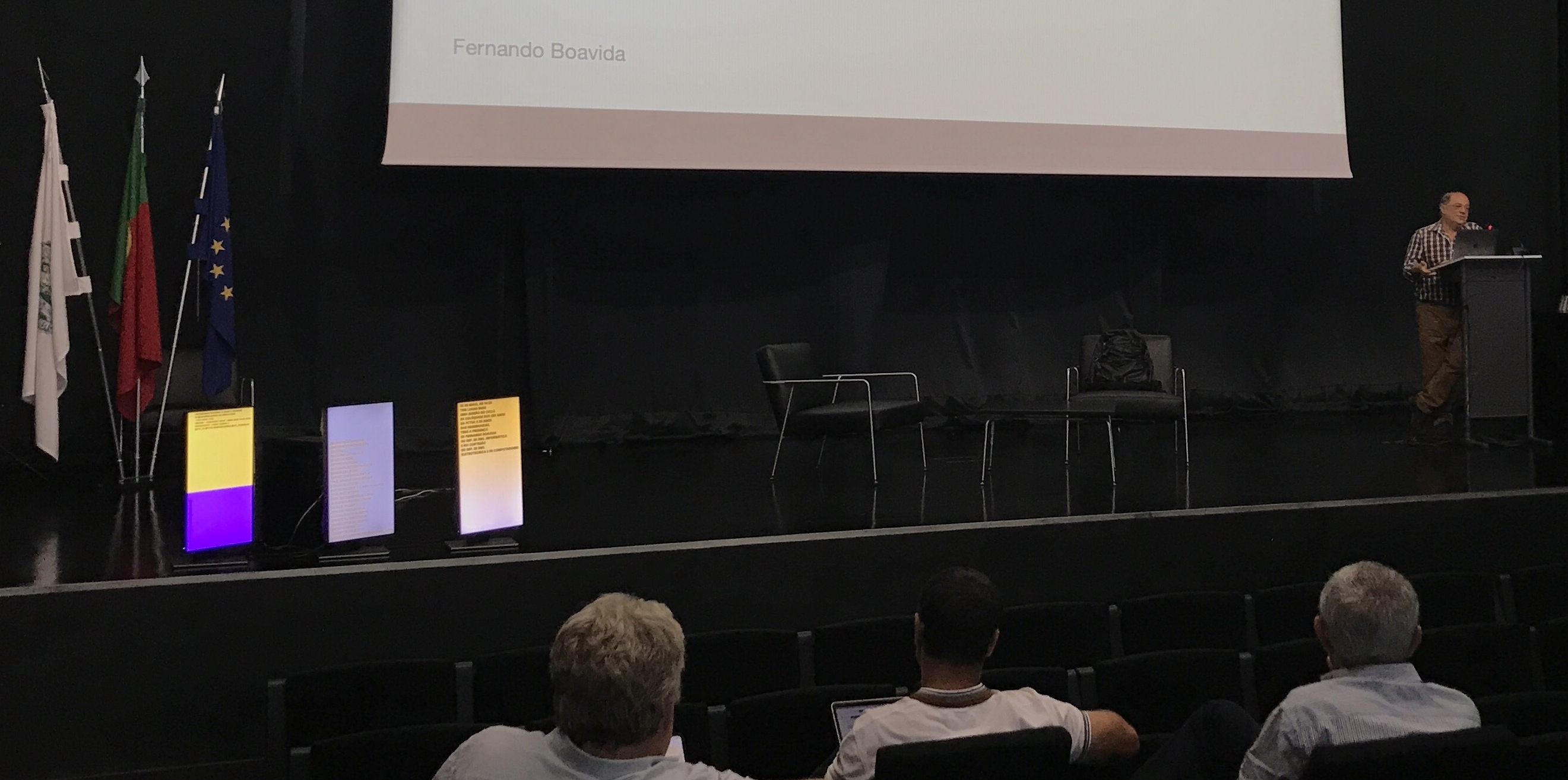}
    \caption{Photography of a new configuration of \esuc~in the commemoration of 250 years of Faculty of Science and Technology of \acs{uc} (May 2022), during one keynote speaker. Photography by Tiago Martins. \textcopyright~2022.}
    \label{fig:pop-up-setup}
\end{figure}

The evaluation experiments showed that some emotions presented on outputs were both recognised by the \acs{ml} classifier and the people. Nevertheless, the presented emotion-audiovisual mapping has still presented some biases (especially cultural-related biases), since it was built over the rules of Western audiovisual rules and presents some cultural limitation. This way, we are aware that the results and conclusions from the emotional perception of the evaluation stage will be not the same in different regions. We will attempt to minimise these biases, in future, through an iterative approach based on the interactive evaluation of outputs followed by the automatic fine-tuning of emotional mapping data. 
Also, we noticed that the automatic translation procedure occasionally leads to some inconsistencies in emotion recognition. This issue is mostly due to that when a sentence is translated to English some words and expressions could be mistranslated or taken out of the context of the original language. Future work on this work will include the use of~\acs{ml} approaches to text translation that also comprised the use of classifiers trained in other languages than English. We also intend to include redundant recognition approaches, such as using lexicon-based approaches, to minimise the current biases. 

We have also the intent of exploring new installation configurations and being able to adapt the installation to other situations and contexts. We have already tested a new physical setup version of the installation, in May 2022, by invitation of the Faculty of Science and Technology of \acs{uc} in the commemoration of its 250 years. Figure \ref{fig:pop-up-setup} documents the presence of the installation at this event.

Future work also will focus on the use of the proposed \acs{cc} approach in the creative commercial scenario. This task will include the 
\rom{1} further study and evaluation of the presented approach in a professional scenario, particularly focusing on the exploratory and co-creativity value of \emph{Typesetter} module as a tool;
\rom{2} further experimentation with digital-based media and formats in order to study the possibility of use this approach to highlight some contents on digital media;
and \rom{3} explore the inclusion of images on the poster designs.

\esuc~promotes a reflection between a city and its inhabitants, the ongoing themes and events that build its ever-changing social portrait, even more mutable when it is displayed through its online trails. Beyond this online portrait, the parallel, physical display embodied in one of its spaces creates a direct confrontation of what Coimbra is in its physical, emotional and cultural dimensions as a unified self, from which we hope its visitors can reflect upon, and even become more active members in their community for change.

\acknowledgments{
\esuc~is an artwork produced with the support of the Institute of Interdisciplinary Research of \acs{uc} (in the scope of the celebration of 730 years of \acs{uc}) and \acs{capc}. We would like to express our gratitude to all the participants in the evaluation sessions. This work is partially supported by Foundation for Science and Technology, I.P./MCTES (Portugal) through national funds (PIDDAC), within the scope of project UIDB/00326/2020. S\'{e}rgio M. Rebelo is funded by FCT under the grant SFRH/BD/132728/2017 and COVID/BD/151969/2021. Mariana Sei\c{c}a is funded by FCT under the grant SFRH/BD/138285/2018.
}

\bibliographystyle{abbrv}

\bibliography{main}
\end{document}